\documentclass{WileyMSP-template}

\usepackage[T1]{fontenc}
\usepackage[utf8]{inputenc}
\usepackage{graphicx}
\usepackage{amsmath,amssymb,bm}
\usepackage{siunitx}
\usepackage[dvipsnames]{xcolor}
\usepackage[hidelinks]{hyperref}
\usepackage[nameinlink,capitalize]{cleveref}
\usepackage{todonotes}
\usepackage{dcolumn}
\usepackage{upgreek}
\usepackage{booktabs}
\usepackage{subcaption}   % for subfigure + \phantomcaption
\usepackage{lipsum}
\usepackage[percent]{overpic}

\usepackage[normalem]{ulem}

\usepackage[backend=biber,
            style=numeric-comp,
            maxnames=6,
            minnames=3,
            doi=true,                      % Shows DOI
            isbn=false,
            url=false,
            giveninits=false,              % Show full given names
            sorting=none]{biblatex}
\graphicspath{{./Figures/}}

\newcommand{\rhoxy}{\rho_{xy}}
\newcommand{\rhoxx}{\rho_{xx}}

\newcommand{\rhoAHE}{\rho_{\mathrm{AHE}}}
\newcommand{\rhoTHE}{\rho_{\mathrm{THE}}}
\newcommand{\sigmaxx}{\sigma_{xx}}
\newcommand{\sigmaAHE}{\sigma_{\mathrm{AHE}}}

\newcommand{\etal}{\emph{et\,al.}}

\DeclareSIUnit{\torr}{torr}
\DeclareSIUnit\angstrom{\text {Å}}
\DeclareSIUnit\ions{ions}

\newcommand*{\Fref}[2][]{%
  \hyperref[{#2}]{%
    Figure~\ref*{#2}% <--- Simply remove the font command
    \ifx\\#1\\%
    \else
      \,#1%
    \fi
  }%
}

\newcommand*{\Eref}[2][]{%
  \hyperref[{#2}]{%
    Equation~\ref*{#2}%
    \ifx\\#1\\%
    \else
      \,#1%
    \fi
  }%
}

\newcommand*{\Aref}[2][]{%
  \hyperref[{#2}]{%
    Appendix~\ref*{#2}%
    \ifx\\#1\\%
    \else
      \,#1%
    \fi
  }%
}

\newcommand*{\Sref}[2][]{%
  \hyperref[{#2}]{%
    Section~\ref*{#2}%
    \ifx\\#1\\%
    \else
      \,#1%
    \fi
  }%
}

\newcommand*{\Tref}[2][]{%
  \hyperref[{#2}]{%
    \bfseries{Table~\ref*{#2}}%
    \ifx\\#1\\%
    \else
      \,#1%
    \fi
  }%
}

\begin{document}

\pagestyle{fancy}
% \rhead{\includegraphics[width=2.5cm]{vch-logo.png}}

\title{Disorder-broadened topological Hall phase and anomalous Hall scaling in FeGe}

\maketitle

% Author: Please give full first and last names for authors and include * after the name of all corresponding authors

\author{Chaman Gupta}
\author{Chris Matsumura}
\author{Hongbin Yang}
\author{Sarah Edwards}
\author{Rebeca M. Gurrola}
\author{Jiun-Haw Chu}
\author{Hanjong Paik}
\author{Yongqiang Wang}
\author{David A. Muller}
\author{Robert Streubel}
\author{Tzu-Ming Lu}
\author{Serena Eley*}

% Dedication

% \dedication{}

% Affiliations: Please provide adacemic titles (Prof. or Dr.) for all authors where applicable, and include an institutional email address for all corresponding authors
\begin{affiliations}
Chaman Gupta\\
Department of Materials Science and Engineering, University of Washington, Seattle,\\Washington 98195, USA\\

\medskip

Chris Matsumura, Sarah Edwards, and Prof. Jiun-Haw Chu\\
Department of Physics, University of Washington, Seattle, Washington 98195, USA\\

\medskip

Dr. Hongbin Yang and Prof. David A. Muller\\
School of Applied and Engineering Physics, Cornell University, Ithaca, New York 14853, USA\\

% \medskip

% Prof. David A. Muller\\
% School of Applied and Engineering Physics, Cornell University, Ithaca, New York 14853, USA\\
\medskip

Dr. Tzu-Ming Lu and Dr. Rebeca M. Gurrola\\
Center for Integrated Nanotechnologies, Sandia National Laboratories, Albuquerque, New Mexico 87123, USA\\
\medskip

Dr. Hanjong Paik\\
School of Electrical and Computer Engineering, University of Oklahoma, Norman, Oklahoma 73019, USA
\medskip

Prof. Robert Streubel\\
Department of Physics and Astronomy, University of Nebraska-Lincoln, Lincoln, NE 68588, USA

\medskip

Dr. Yongqiang Wang\\
Center for Integrated Nanotechnologies, Los Alamos National Laboratory, Los Alamos, NM 87545 USA

\medskip

Prof. Serena Eley\\
Department of Electrical and Computer Engineering, University of Washington, Seattle, Washington 98195, USA\\
Department of Physics, Colorado School of Mines, Golden, Colorado 80401, USA\\
Email Address: serename@uw.edu\\

\end{affiliations}

% Keywords: Please provide a minimum of three and a maximum of seven keywords, separated by commas


% Abstract should be written in the present tense and impersonal style (i.e., avoid we), and be at most 200 words long
\begin{abstract}

Magnetic skyrmions are topologically protected spin textures that are promising candidates for low-power spintronic memory and logic devices. Realizing skyrmion-based devices requires an understanding of how structural disorder affects their stability and transport properties. This study uses Ne$^{+}$ ion irradiation at fluences from $10^{11}$ to $10^{14}$ ions-cm$^{-2}$ to systematically vary defect densities in \SI{80}{\nm} epitaxial FeGe films and quantify the resulting modifications to magnetic phase boundaries and electronic scattering. Temperature- and field-dependent Hall measurements reveal that increasing disorder progressively extends the topological Hall signal from a narrow window near \SI{200}{\kelvin} in pristine films down to \SI{4}{\kelvin} at the highest fluence, with peak amplitude more than doubling. Simultaneously, the anomalous Hall effect transitions from quadratic Berry curvature scaling to linear skew scattering behavior, with the skew coefficient increasing threefold. These results establish quantitative correlations between defect concentration, skyrmion phase space, and transport mechanisms in a chiral magnet. It demonstrates that ion-beam modification provides systematic control over both topological texture stability and electrical detectability.

\end{abstract}

% Text: Please use section headings and subheadings as specified below. For communications, all section headings apart from Experimental Section should be removed
% Please make the first reference to a display item bold: \textbf{Figure 1}
% Do not abbreviate Figure, Equation, etc.; display items are always singular, i.e., Figure 1 and 2.
% Equations are always singular, i.e., Equation 1 and 2, and should be inserted using the {equation} environment, not as graphics
% Please do not use footnotes in the text, additional information can be added to the Reference list.

% =========================
% INTRODUCTION
% =========================

\section{Introduction}\label{sec:intro}

Magnetic skyrmions are nanometer-scale, topologically non-trivial spin textures stabilized by a competition between ferromagnetic exchange and antisymmetric Dzyaloshinskii-Moriya interactions (DMI)~\cite{skyrmeUnifiedFieldTheory1962, bogdanovThermodynamicallyStableMagnetic1994, muhlbauerSkyrmionLatticeChiral2009a, neubauerTopologicalHallEffect2009}. In non-centrosymmetric magnets, the DMI introduces a chiral term, $H_{\textrm{DMI}} = -\sum_{i,j}\mathbf{D}_{ij}\cdot(\mathbf{S}_i\times\mathbf{S}_j)$, where $\mathbf{D}_{ij}$ is the DMI vector between sites $i$ and $j$, and  $\mathbf{S}_i$ denotes the spin at site $i$~\cite{dzyaloshinskyThermodynamicTheoryWeak1958, moriyaAnisotropicSuperexchangeInteraction1960}. This interaction energetically favors the twisted configurations over collinear ferromagnetic order~\cite{nagaosaTopologicalPropertiesDynamics2013}. When sufficiently strong, the DMI drives the formation of helical, conical, and skyrmionic spin phases~\cite{higginsSignatureHelimagneticOrdering2010, dussauxLocalDynamicsTopological2016}. The skyrmion's topological charge, $Q=\frac{1}{4\pi}\int m(\mathbf{r})\cdot[\partial_x m(\mathbf{r})\times\partial_y m(\mathbf{r})]\,\mathrm{d}x\,\mathrm{d}y$, where $\textbf{m(r)}$ is the normalized magnetization field, confers an energy barrier against unwinding into the ferromagnetic state~\cite{heinzeSpontaneousAtomicscaleMagnetic2011, raftreyQuantifyingTopologyMagnetic2024}. Skyrmions can be displaced by applied electrical currents through spin-transfer torque, where the critical current density, $j_c$, is defined as the minimum current required to induce continuous skyrmion motion~\cite{jonietzSpinTransferTorques2010}. Experiments and simulations demonstrate that skyrmions exhibit \SIrange[scientific-notation = true, print-unity-mantissa = false]{1e6}{1e7}{\A\per\square\meter}~\cite{yuSkyrmionFlowRoom2012, okuyamaDeformationMovingMagnetic2019, iwasakiUniversalCurrentvelocityRelation2013}, three to six orders of magnitude lower than the currents required to move magnetic domain walls \cite{jiangEnhancedStochasticityDomain2010}. The topological protection, combined with nanoscale dimensions and ultralow critical current density, makes skyrmions promising candidates for information storage and logic in spintronic devices~\cite{jiangDirectObservationSkyrmion2017, wooCurrentdrivenDynamicsInhibition2018, fertMagneticSkyrmionsAdvances2017, zhangMagneticSkyrmionsMaterials2023}.

\medskip

While skyrmions are often investigated in ideal thin films, translating these findings into functional devices requires understanding how structural disorder affects their stability and dynamics. Defects such as vacancies, interstitials, or grain boundaries create spatially varying potentials that may pin~\cite{limafernandesUniversalityDefectskyrmionInteraction2018}, distort~\cite{derras-choukSkyrmionsDefects2021}, or annihilate skyrmions~\cite{iwasakiCurrentinducedSkyrmionDynamics2013}, thereby modifying current-driven trajectories and electrical transport signatures~\cite{reichhardtStaticsDynamicsSkyrmions2022, linParticleModelSkyrmions2013}. A few experimental studies suggest that  pinning may suppress the skyrmion Hall effect—transverse deflection that complicates racetrack geometries—while potentially enhancing thermal stability~\cite{derras-choukSkyrmionsDefects2021, chenSkyrmionHallEffect2017}. Conversely, excessive disorder may immobilize or destroy skyrmions~\cite{feilhauerControlledMotionSkyrmions2020}. Establishing relationships between defect density, magnetic phase boundaries, and transport properties remains a central challenge in skyrmion device engineering.

\medskip

We address this challenge with two objectives: first, to determine how atomic-scale disorder controls skyrmion formation and low-temperature stability; second, to identify defect-engineering strategies that beneficially manipulate skyrmion dynamics for device applications. To tune the disorder landscape, we ion-beam modify epitaxial FeGe films, systematically varying defect density over three orders of magnitude. Electrical transport and magnetization measurements reveal two key findings. Firstly, in pristine FeGe films, skyrmions are stable over a limited temperature range -- from approximately \SI{280}{\kelvin} down to \SI{80}{\kelvin}. As disorder is introduced and increased, this lower bound decreases: with the highest defect concentration, skyrmion stability extends all the way down to \SI{4}{\kelvin}. Thus, increasing disorder significantly broadens the temperature range over which skyrmions can exist. Secondly, the same defects that stabilize skyrmions also modify the scattering of charge carriers, shifting the dominant electronic transport mechanism. Together, these results provide design rules linking defect density to both skyrmion stability and electrical detectability, establishing a framework for disorder-optimized spintronic devices.

% =========================
% RESULTS & DISCUSSION
% =========================
\section{Results and Discussion}\label{sec:results}

\subsection{Sample preparation and disorder characterization}
\label{sec:results_srim}
% \lipsum[1-2]

Among B20-type chiral magnets~\cite{yuRoomtemperatureFormationSkyrmion2011, yuRealspaceObservationTwodimensional2010}, FeGe has emerged as a prototypical system for studying skyrmion formation and stability~\cite{huangExtendedSkyrmionPhase2012, bocarslyMagnetoentropicSignaturesSkyrmionic2018, gallagherRobustZeroFieldSkyrmion2017}. Its relatively high Curie temperature ($T_c \approx \SI{278}{\kelvin}$)~\cite{haraldsonMagneticResonanceCubic1972} and moderate DMI strength ($D \approx \SI{1.6}{\mJ\per\square\m}$)~\cite{turgutEngineeringDzyaloshinskiiMoriyaInteraction2018} produce a helical ground state with a characteristic wavelength of $\lambda \simeq \SI{70}{\nm}$~\cite{yuRoomtemperatureFormationSkyrmion2011}. When grown as thin epitaxial films on Si(111) substrates, strain-induced uniaxial anisotropy markedly broadens the skyrmion stability window~\cite{budhathokiRoomtemperatureSkyrmionsStrainengineered2020,huangExtendedSkyrmionPhase2012}. FeGe(111) thin-films of \SI{75}{\nm} sustain ordered skyrmion lattices from near $T_c$ down to $\sim \SI{200}{\kelvin}$~\cite{yuRoomtemperatureFormationSkyrmion2011}, and zero-field skyrmions appear when interfacial anisotropy compensates the demagnetizing field~\cite{gallagherRobustZeroFieldSkyrmion2017}. At zero field and intermediate temperatures, FeGe adopts a helical ground state in which the magnetization rotates periodically along a single crystallographic direction with a period of approximately \SI{70}{\nm}, forming a one-dimensional spin spiral~\cite{lebechMagneticStructuresCubic1989}. As the magnetic field is applied perpendicular to the film plane, this helical order transforms into a conical phase where spins cant uniformly toward the field direction while maintaining a spiraling component~\cite{muhlbauerSkyrmionLatticeChiral2009a}. Within a narrow field and temperature window between the helical and conical phases, a hexagonal lattice of skyrmions becomes energetically favorable~\cite{yuRoomtemperatureFormationSkyrmion2011, huangExtendedSkyrmionPhase2012}. Each skyrmion consists of spins that rotate from pointing antiparallel to the field at the core, to parallel at the periphery, creating a localized vortex-like structure~\cite{nagaosaTopologicalPropertiesDynamics2013}. 

\begin{figure*}[!ht]
    \centering
    \begin{subcaptiongroup}
        \phantomcaption\label{fig:image:SRIM}
        \phantomcaption\label{fig:image:MAADFS2}
        \phantomcaption\label{fig:image:MAADFS5}
        \phantomcaption\label{fig:image:MEPD1}
        \phantomcaption\label{fig:image:MEPD2}
        \phantomcaption\label{fig:image:MEPD3}
    \end{subcaptiongroup}
    \includegraphics[width=\textwidth]{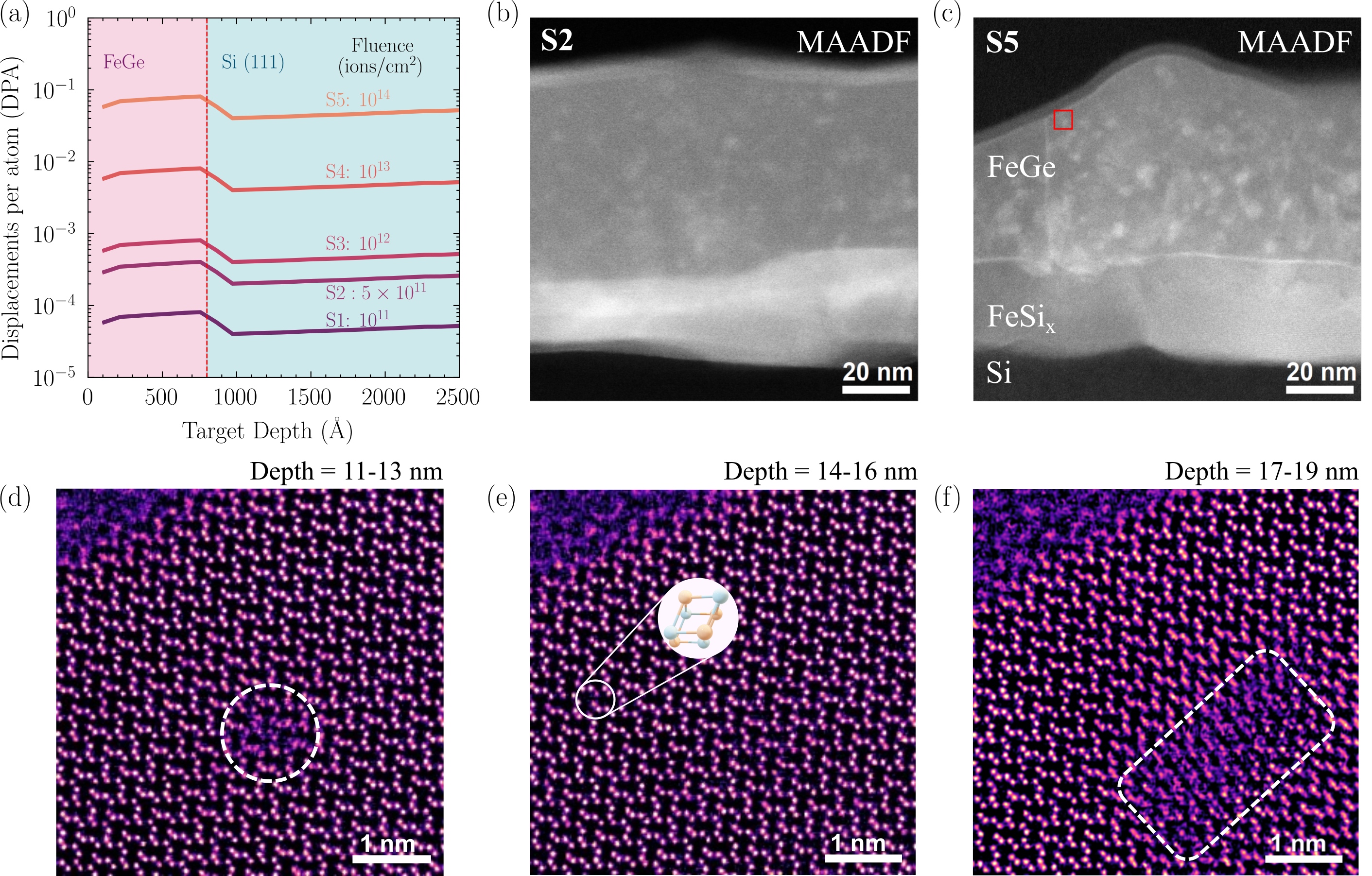}
    \caption{\textbf{Identification of defects.} \protect\subref{fig:image:SRIM} SRIM full-cascade calculations for \SI{400}{\keV} Ne$^{+}$ irradiation (fluences $10^{11}–10^{14}$~ions-cm$^{-2}$; samples S1--S5) give depth-resolved displacements-per-atom (dpa). The FeGe film (0--\SI{80}{\nm}, pink) experiences a nearly uniform damage level. \protect\subref{fig:image:MAADFS2} Cross-sectional medium angle annular dark-field (MAADF) STEM of a lightly irradiated film (S2, $5\times10^{11}$~ions-cm$^{-2}$), and, \protect\subref{fig:image:MAADFS5} of the highest-dose fillm (S5, $10^{14}$~ions-cm$^{-2}$) reveal an increasing density of nanoscale contrast variations attributed to point-defect clusters within the FeGe layer. Atomic-resolution multislice electron ptychography images taken from the red boxed area in \protect\subref{fig:image:MAADFS5} for depths of \protect\subref{fig:image:MEPD1}\SIrange{11}{13}{\nm}, \protect\subref{fig:image:MEPD2}\SIrange{14}{16}{\nm}, and \protect\subref{fig:image:MEPD3}\SIrange{17}{19}{\nm}}
    \label{fig:srimstem}
\end{figure*}

\medskip

In this work, epitaxial FeGe films with a nominal thickness of \SI{300}{\nm} were grown on Si(111) substrates by molecular beam epitaxy following a process used to grow epitaxial $\mathrm{Mn}_{x}\mathrm{Fe}_{1-x}\mathrm{Ge}$ films by Turgut \etal~\cite{turgutEngineeringDzyaloshinskiiMoriyaInteraction2018, venutiInducingTunableSkyrmionantiskyrmion2024, liuStructuralPropertiesRecrystallization2025}, and milled down to \SI{80}{\nm} (see Methods~\Sref{sec:methods:irr} for growth details). Each wafer was diced into smaller chips: \qtyproduct{4 x 4}{\milli\metre}
 pieces for magnetization measurements and \qtyproduct{6 x 6}{\milli\metre}
 pieces for transport studies. Six-terminal Hall bar devices were fabricated on the larger chips using standard photolithography and Ar$^{+}$-ion milling. Each chip contained multiple Hall bars with typical channel dimensions of \SI{150}{\um} (width) and \SI{800}{\um} (length) between voltage contacts (see Methods~\Sref{sec:methods:irr} and Supplementary Figure~S1 for device geometry). To introduce controlled structural disorder, we irradiated chips with $\mathrm{Ne}^{+}$ ions at \SI{400}{\keV} incident with five different fluences. One sample was retained as an unirradiated reference. Sample ID and corresponding fluence is provided in~\textbf{\Fref{fig:image:SRIM}}.
% The ion beam was placed over the sample surface to ensure uniform exposure, and
The chamber was maintained at base pressure below \SI[scientific-notation = true, print-unity-mantissa = false]{1e-6}{\torr} throughout irradiation (see Methods \Sref{sec:methods:irr} for more details regarding the irradiation protocol).

\medskip

We directly observe defects in the irradiated FeGe films by scanning transmission electron microscope (STEM) imaging. As shown in~\Fref{fig:image:MAADFS2} and~\ref{fig:image:MAADFS5}, irradiation-induced defects appear as bright dots in the medium angle annular dark field (MAADF) images. In sample S2 (\Fref{fig:image:MAADFS2}), the distance between defects is on the order of \SI{10}{\nm}. On the other hand, in sample S5 (\Fref{fig:image:MAADFS5}), we observe a denser defect distribution, with larger bright spots likely due to multiple defects overlapping along the electron beam direction. To resolve the defect atomic structure and distribution, multislice electron ptychography (MEP) was performed and is shown in~\Fref{fig:image:MEPD1}--\ref{fig:image:MEPD3}. We focus on a region near the surface of S5 (red rectangle in~\Fref{fig:image:MAADFS5}), where we see two defects with varying sizes, as shown in~\Fref{fig:image:MEPD1} and~\ref{fig:image:MEPD3} at different depths. They are separated by about \SI{3}{\nm}. The region between these two defects is nearly pristine observed in the~\Fref{fig:image:MEPD2} along with a unit cell schematic shown with in the blown-up inset. This observation demonstrates that both defect density and size distribution increase with increase in irradiation fluence. Next, we study the effects of the varying defect density on magneto-transport measurements in the FeGe.

\medskip

\subsection{Topological Hall Effect (THE)}
\label{sec:results:the}

Skyrmions produce a measurable electrical signature through the topological Hall effect (THE)~\cite{nagaosaTopologicalPropertiesDynamics2013}. As charge carriers traverse a skyrmion texture, they acquire a Berry phase from the non-collinear spin arrangement, which acts as an effective magnetic field that deflects their trajectories perpendicular to the applied current~\cite{brunoTopologicalHallEffect2004a, neubauerTopologicalHallEffect2009}. This deflection generates a transverse voltage, THE, that serves as a transport proxy for skyrmion density~\cite{nagaosaTopologicalPropertiesDynamics2013}. Although THE signals can arise from other non-collinear magnetic structures such as spin spirals or magnetic bubbles~\cite{kimbellChallengesIdentifyingChiral2022}, skyrmions have been directly observed in FeGe films via Lorentz transmission electron microscopy (LTEM)\cite{yuRoomtemperatureFormationSkyrmion2011, yuSkyrmionFlowRoom2012}, and THE has been established as a robust signature of the skyrmion phase in this material \cite{gallagherRobustZeroFieldSkyrmion2017, huangExtendedSkyrmionPhase2012}.

\begin{figure*}[!t]
    \centering
    \begin{subcaptiongroup}
        \phantomcaption\label{fig:the:extraction:S0}
        \phantomcaption\label{fig:the:extraction:S5}
        \phantomcaption\label{fig:the:temperature:S0}
        \phantomcaption\label{fig:the:temperature:S5}
    \end{subcaptiongroup}
    \includegraphics[width=\textwidth]{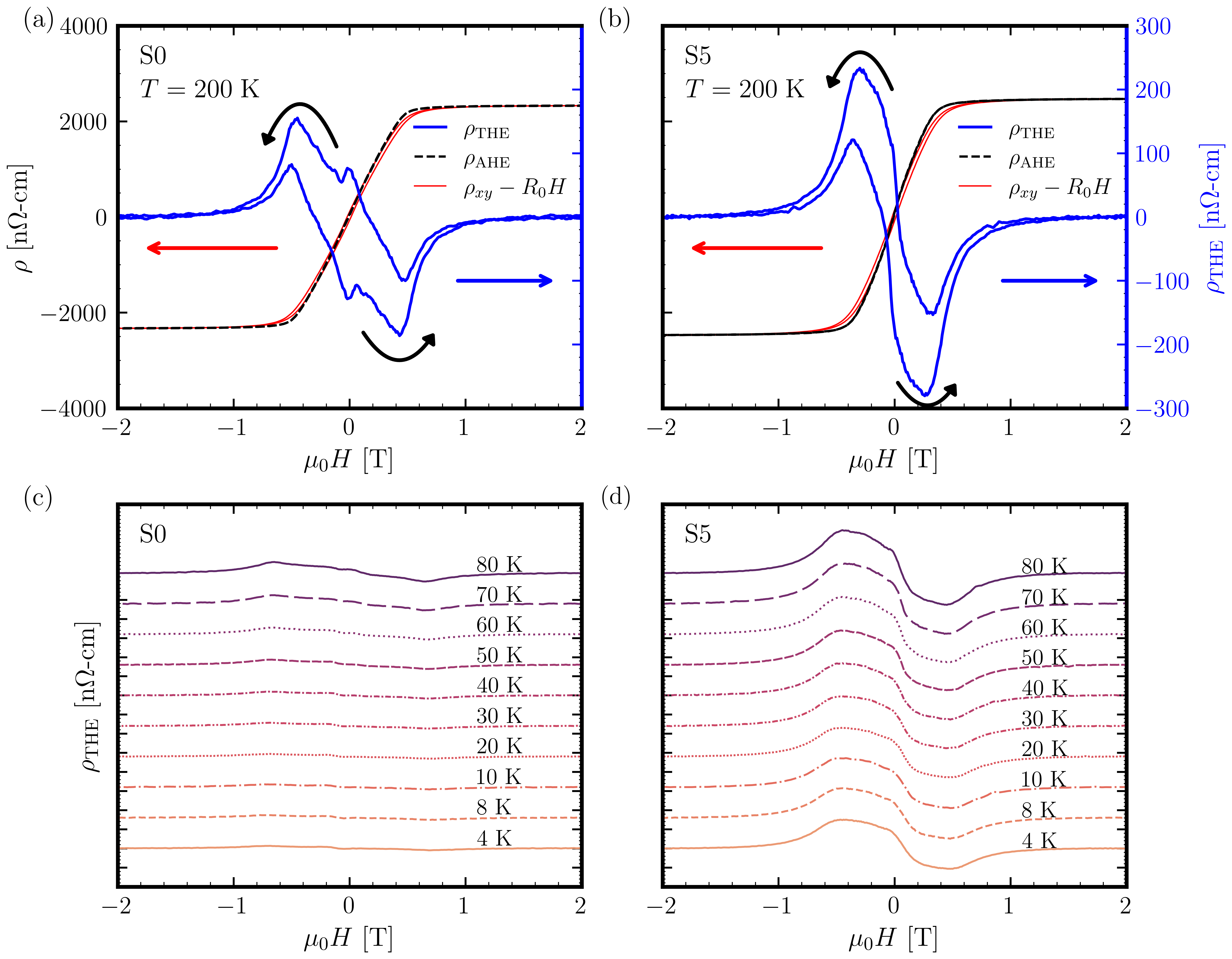}
    \caption{Topological Hall extraction and temperature evolution in pristine and irradiated FeGe. \protect\subref{fig:the:extraction:S0} Decomposition of the Hall signal at $T = 200~\mathrm{K}$ for pristine S0 over the field range $-2 \le \mu_0H \le 2~\mathrm{T}$. Total Hall resistivity after ordinary Hall removal, $\rho_{xy} - R_0H$ (red, left $y$-axis), anomalous Hall contribution $\rho_{\mathrm{AHE}}$ (black dashed, left $y$-axis), and residual topological Hall resistivity $\rho_{\mathrm{THE}}$ (blue, right $y$-axis). Black curved arrows indicate field-sweep direction; colored arrows mark corresponding $y$-axes. \protect\subref{fig:the:extraction:S5} Same decomposition for highly disordered S5 ($10^{14}~\mathrm{ions\,cm}^{-2}$), showing enhanced THE amplitude compared to pristine film. \protect\subref{fig:the:temperature:S0} Temperature evolution of $\rho_{\mathrm{THE}}(H)$ from $4~\mathrm{K}$ to $80~\mathrm{K}$ in pristine S0 (curves vertically offset for clarity), revealing diminishing THE signals at low temperatures. \protect\subref{fig:the:temperature:S5} Temperature evolution of $\rho_{\mathrm{THE}}(H)$ in irradiated S5, showing persistent THE signals across all temperatures, demonstrating disorder-stabilized skyrmions at low temperatures.}
    \label{fig:the-extraction}
\end{figure*}

\medskip

We measured the total Hall resistivity in devices containing six Hall bars per chip using an Oxford Teslatron cryostat with out-of-plane magnetic fields up to $\pm \SI{3}{\tesla}$ and temperatures from \SIrange{4}{300}{\kelvin}. Longitudinal and transverse voltages were acquired simultaneously using standard low-noise AC lock-in techniques with applied currents of approximately \SI{20}{\uA}, with further details provided in the Methods~\Sref{sec:methods:transport}. Supplementary Figure~S1 shows the device geometry and wiring schematic. The total transverse resistivity, $\rhoxy(H, T)$, measured in a ferromagnet that hosts nontrivial spin textures, contains three contributions which must be separated to isolate the topological Hall component~\cite{addisonAnomalousTopologicalHall2024}:

\begin{equation}
    \rhoxy(H,T) = \rho_{OHE}(H) + \rhoAHE\bigl(M(H,T)\bigr) + \rhoTHE(H,T),
\label{eq:rho_xy}
\end{equation}

where $\rho_{OHE}$ is the ordinary Hall component, $H$ is the applied magnetic field, $\rho_{\mathrm{AHE}}$ is the anomalous Hall resistivity that scales with magnetization $M$~\cite{nagaosaAnomalousHallEffect2010}, and $\rho_{\mathrm{THE}}$ is the topological Hall resistivity from skyrmions~\cite{nagaosaTopologicalPropertiesDynamics2013}. The ordinary Hall effect arises from the Lorentz force on charge carriers, while the anomalous Hall effect originates from spin-orbit coupling in the ferromagnetic state~\cite{nagaosaAnomalousHallEffect2010, nagaosaTopologicalPropertiesDynamics2013}.

\medskip

\textbf{\Fref{fig:the-extraction}} demonstrates this extraction of $\rhoTHE$ and its temperature evolution for pristine (S0) and most disordered (S5) samples. \Fref{fig:the:extraction:S0} and~\ref{fig:the:extraction:S5} show the decomposition procedure at $T=\SI{200}{K}$, where the skyrmion phase is established in FeGe films~\cite{yuRoomtemperatureFormationSkyrmion2011}. We first subtract the linear ordinary Hall background, $R_0H$, where $R_0$ is determined from the high-field slope (solid red curves, left axes). In the saturated high-field region ($\mu_0H > \SI{1.5}{\tesla}$), all topological textures collapse into the ferromagnetic state and only the anomalous Hall effect remains. We fit this region to obtain $\rhoAHE(M)$ (black-dashed line), using magnetization data measured via SQUID magnetometry (see Methods \Sref{sec:methods:mpms}). The residual signal after subtracting both ordinary and anomalous contributions yields the\\$\rhoTHE(H)$ (blue curves, right axes). Two pronounced peaks of opposite sign appear symmetrically about zero field, reaching amplitudes near $\pm\SI{150}{\nano\ohm\cm}$. In FeGe, the skyrmion phases emerge from a helical ground state that transforms into a conical phase under applied field, before fully saturating into the ferromagnetic state at higher fields~\cite{liHighFieldMagnetic2023}. The THE peaks occur near the field range where the skyrmion lattice is most stable, between the helical-to-conical and conical-to-ferromagnetic transitions~\cite{chenTopologicalHallEffect2025}. Small hysteresis between field-sweep directions reflects metastable domain configurations during skyrmion nucleation and annihilation~\cite{akyildizSkyrmionsDynamicMagnetic2025}.

\medskip

The temperature evolution of $\rhoTHE(H)$ reveals a dramatic effect of disorder on skyrmion stability. In pristine FeGe, S0 (\Fref{fig:the:temperature:S0}), the THE signal decreases systematically from \SI{80}{\kelvin}, with the peak amplitudes dropping from \SI{80}{\nano\ohm\cm} at \SI{80}{\kelvin} to barely detectable levels by \SI{50}{\kelvin}. By \SI{4}{\kelvin}, the THE is essentially absent, indicating that skyrmions become thermodynamically unfavorable at low temperatures in clean FeGe films. In a stark contrast, the most disordered sample, S5 (\Fref{fig:the:temperature:S5}, exhibits robust THE signals across the entire temperature ranges from \SIrange{4}{80}{\kelvin} (curves offset vertically for clarity). Peak amplitudes reach approximately \SI{150}{\nano\ohm\cm} at \SI{80}{\kelvin} and remain near \SI{80}{\nano\ohm\cm} even at \SI{4}{\kelvin}. Moreover, the field width over which the $\rhoTHE$ remains finite broadens at all temperatures in S5 compared to S0. The persistence of large THE signals down to \SI{4}{\kelvin} in irradiated thin-films of \SI{80}{\nm} demonstrates that atomic-scale disorder plays a strong role in stabilizing skyrmions at low temperatures where they would otherwise be thermodynamically unfavorable~\cite{huangExtendedSkyrmionPhase2012}.

\begin{figure*}[!t]
    \centering
    \begin{subcaptiongroup}
        \phantomcaption\label{fig:the:all}
        \phantomcaption\label{fig:the:S0}
        \phantomcaption\label{fig:the:S1}
        \phantomcaption\label{fig:the:S3}
        \phantomcaption\label{fig:the:S4}
        \phantomcaption\label{fig:the:S5}
    \end{subcaptiongroup}

    \includegraphics[width=\textwidth]{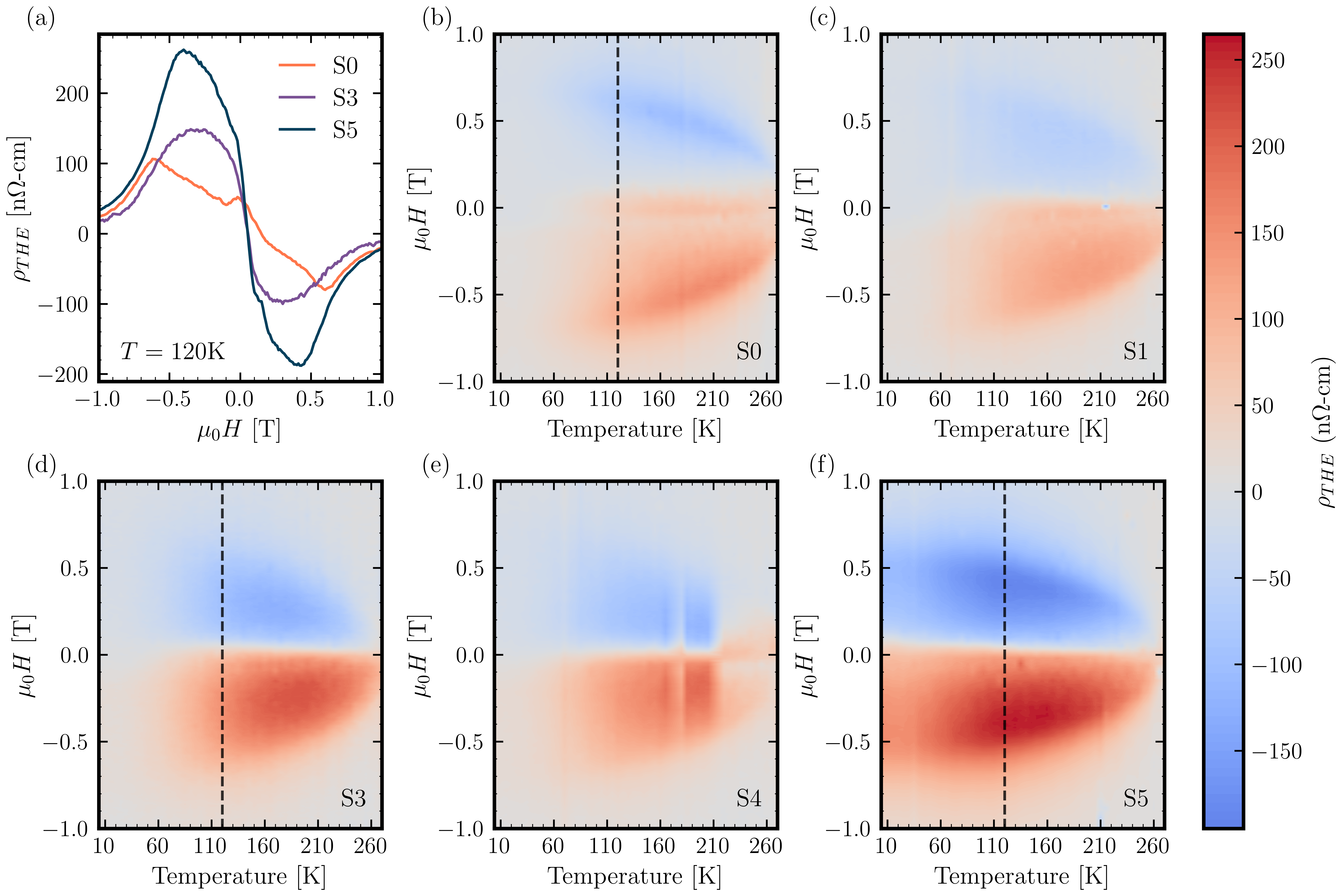}
    \caption{\textbf{Disorder evolution of the topological Hall effect.}
    \protect\subref{fig:the:all} Field dependence of the topological Hall resistivity
    $\rho_{\mathrm{THE}}(H)$ at $T=120$\,K for S0 (pristine), S3, and S5 over the range
    $-1\le \mu_0H \le 1$\,T, showing a progressive increase in amplitude and width with irradiation fluence.
    Temperature–field maps of $\rho_{\mathrm{THE}}(T,H)$ for
    \protect\subref{fig:the:S0} S0,
    \protect\subref{fig:the:S1} S1 ($10^{11}$\,ions/cm$^{2}$),
    \protect\subref{fig:the:S3} S3 ($10^{12}$\,ions/cm$^{2}$),
    \protect\subref{fig:the:S4} S4 ($10^{13}$\,ions/cm$^{2}$), and
    \protect\subref{fig:the:S5} S5 ($10^{14}$\,ions/cm$^{2}$)
    are plotted with identical axes and a common color scale (right; red/blue denote positive/negative $\rho_{\mathrm{THE}}$).
    The vertical dashed lines in \protect\subref{fig:the:S0}, \protect\subref{fig:the:S3}, and \protect\subref{fig:the:S5} mark $T=\SI{120}{K}$, corresponding to the cut shown in \protect\subref{fig:the:all}.
    Together, the maps reveal a systematic broadening and strengthening of the low-temperature $\rho_{\mathrm{THE}}$ lobe with increasing defect density.}
    \label{fig:the-maps}
\end{figure*}

\medskip

\textbf{\Fref{fig:the-maps}} illustrates how disorder modifies the skyrmion phase boundaries. Specifically,~\Fref{fig:the:all} compares $\rhoTHE(H)$ at $T=$ \SI{120}{\kelvin} for pristine (S0), and two irradiated films (S3 and S5). Peak THE amplitude increases from $\sim$\SI{100}{\nano\ohm\cm} (S0) to \SI{230}{\nano\ohm\cm} (S5), while the field range broadens from \SIrange{\pm 0.5}{\pm1}{\tesla}. This is consistent with theoretical expectations that defects may enhance the collective pinning landscape and stabilize skyrmions to higher fields~\cite{reichhardtStaticsDynamicsSkyrmions2022, hendersonSkyrmionAlignmentPinning2022}.

\medskip

\Fref{fig:the:S0}--\ref{fig:the:S5} display the temperature-field maps of $\rhoTHE(H,T)$ for all samples with identical axes and color scale (red/blue $\mapsto$ positive/negative $\rhoTHE$) obtained when sweeping the field down from \SIrange{+3}{-3}{\tesla}. The sign convention is preserved across all fluences, confirming that disorder does not alter the chiral handedness imposed by the DMI~\cite{muhlbauerSkyrmionLatticeChiral2009}. In pristine FeGe (\Fref{fig:the:S0}), clear THE signals appear only above $\sim\SI{80}{\kelvin}$ and vanish by $\SI{50}{\kelvin}$, consistent with reports on epitaxial FeGe samples of similar-thickness ($\SIrange{50}{80}{\nm}$) ~\cite{huangExtendedSkyrmionPhase2012, ahmedChiralBobbersSkyrmions2018, gallagherRobustZeroFieldSkyrmion2017, budhathokiRoomtemperatureSkyrmionsStrainengineered2020}. At low fluence (S1, \Fref{fig:the:S1}), the phase boundaries shift minimally. However, at \SI[scientific-notation = true, print-unity-mantissa = false]{1e-12}{\ions\per\square\cm}, a distinct low-temperature THE region emerges and extendes to \SI{30}{\kelvin}. This region grows systematically at higher fluences: S4 (\Fref{fig:the:S4}), shows THE across temperatures, and S5 (\Fref{fig:the:S5}), exhibits the strongest low-temperature THE, with amplitudes much closer to the high-temperature values.

\medskip

Theoretical models have predicted that defects may create energy minima that trap nucleated skyrmions, preventing their collapse into competing magnetic phases~\cite{limafernandesUniversalityDefectskyrmionInteraction2018, reichhardtStaticsDynamicsSkyrmions2022, reichhardtPlasticFlowSkyrmion2020}. Previous simulations by Hoshino et al.\ show that disorder can extend skyrmion stability below equilibrium phase boundaries through pinning-induced metastability, analogous to vortex-glass phases in superconductors~\cite{hoshinoTheoryMagneticSkyrmion2018}. Experiments on chemically disordered $\textrm{MnSi}_{1-x}\textrm{Ga}_x$ and $\textrm{Co}_8\textrm{Zn}_8\textrm{Mn}_4$ similarly demonstrate that disorder extends skyrmion stability below the equilibrium phase boundary~\cite{karubeDisorderedSkyrmionPhase2018}.

\medskip

The increased THE amplitude at high disorder also points to an enhanced skyrmion density. Since, $\rhoTHE \propto n_{\textrm{sk}}$, where $n_{\textrm{sk}}$ is the skyrmion density, the factor-of-two amplitude increase in S5 relative to S0 suggests doubled-skyrmion populations. Assuming the topological charge per skyrmion remains fixed by the DMI, defects likely act as nucleation sites that increase skyrmion density. Ion irradiation can therefore provide a systematic method to extend skyrmion stability over temperature and field ranges relevant for devices~\cite{venutiInducingTunableSkyrmionantiskyrmion2024, liuStructuralPropertiesRecrystallization2025}. However, pinning may impede current-driven motion, motivating analysis of how disorder modifies electronic scattering~\cite{koshibaeTheoryCurrentdrivenSkyrmions2018, akyildizSkyrmionsDynamicMagnetic2025}. In the next section, we examine the anomalous Hall effect to quantify changes in transport mechanisms across the same disorder series.

\medskip

\subsection{Anomalous Hall Effect (AHE)}
\label{sec:ahe-overview}

\begin{figure*}[!t]
    \centering
    \begin{subcaptiongroup}
        \phantomcaption\label{fig:ahe:temperature}
        \phantomcaption\label{fig:ahe:sigmaxx}
    \end{subcaptiongroup}
\includegraphics[width=0.95\textwidth]{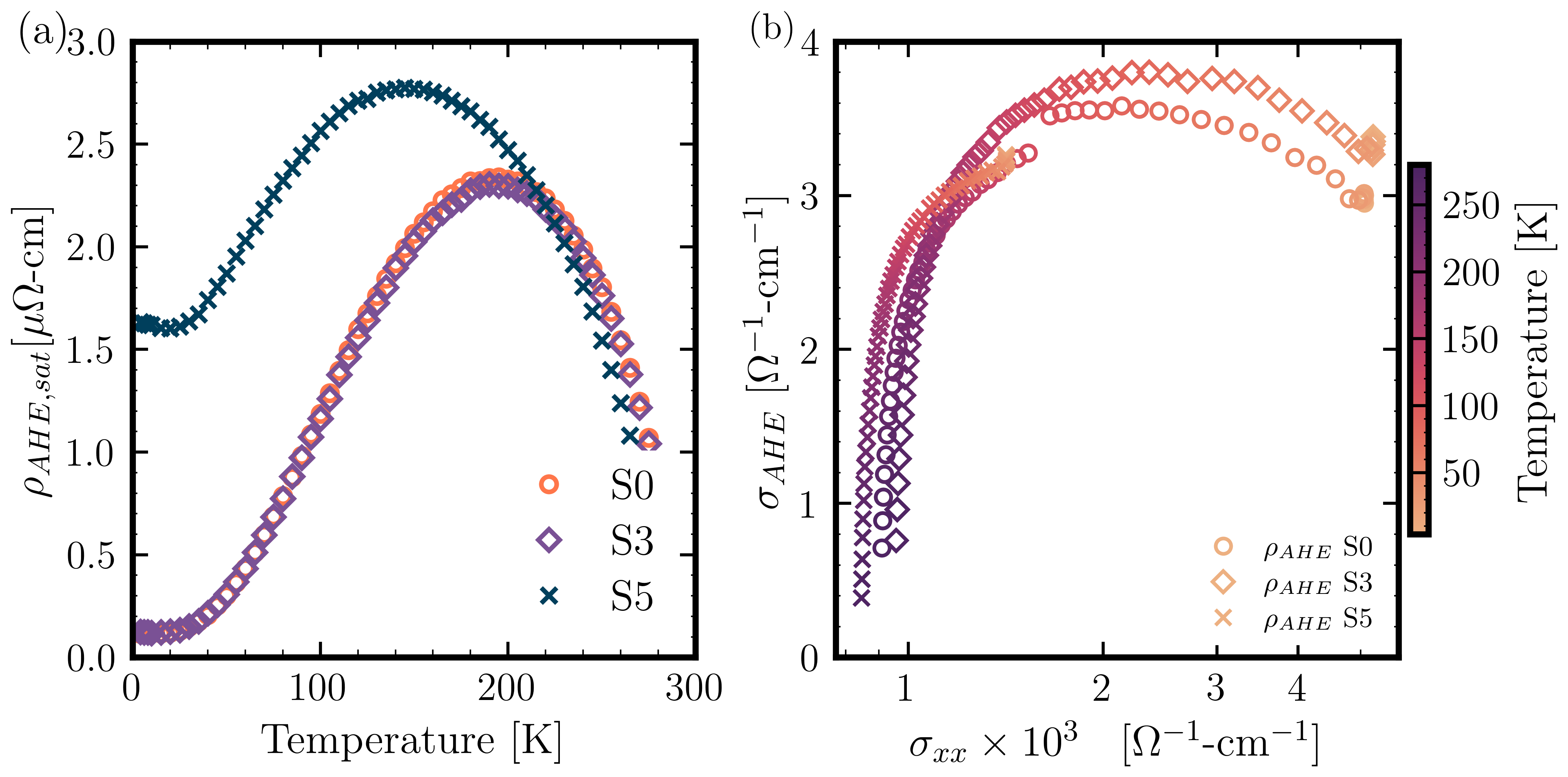}
\caption{\textbf{Anomalous Hall effect across disorder.}
\protect\subref{fig:ahe:temperature} Saturated anomalous Hall resistivity $\rho_{\mathrm{AHE,sat}}(T)$ for S0, S3, and S5, extracted from the positive-field saturation window ($\mu_0H>\SI{1.5}{T}$). Disorder enhances the low-$T$ AHE and slightly shifts the peak near \SIrange{150}{200}{K}. \protect\subref{fig:ahe:sigmaxx} $\sigmaAHE$ vs.\ $\sigmaxx$ with temperature as a parametric color (\SIrange{10}{280}{K}), mapping the conductivity regimes: trajectories move from the high-$\sigmaxx$ intrinsic plateau at high $T$ toward extrinsic, disorder-dominated behavior at lower $\sigmaxx$; irradiation shifts the curves to smaller $\sigmaxx$.}
\label{fig:ahe_overview}
\end{figure*}

Beyond the topological Hall signal from skyrmions, the saturated high-field Hall resistivity, $\rhoAHE$, provides a window into how disorder modifies electronic scattering. In ferromagnets, the anomalous Hall effect (AHE) produces transverse voltages that scale with magnetization~\cite{nagaosaAnomalousHallEffect2010}. This arises from the interplay of spin-orbit interactions and ferromagnetic order, and can exceed the ordinary Hall signal by orders of magnitude. $\rhoAHE$ is measured as the transverse voltage normalized by current and sample geometry, extracted from the high-field saturated region where all magnetic domains align. The AHE has three physical origins. First, the intrinsic mechanism, which arises from Berry curvature in the electronic band structure. Electrons moving through the momentum space accumulate a transverse displacement independent of scattering events~\cite{vermaUnifiedTheoryAnomalous2022}. The other two effects are extrinsic in nature. Specifically, extrinsic skew-scattering, where spin-orbit-coupled impurities deflect carriers asymmetrically, producing a Hall signal proportional to the number of scattering events~\cite{smitSpontaneousHallEffect1958}. Lastly, extrinsic side-jump scattering, where each collision displaces the carrier laterally by a fixed amount~\cite{smitSideJumpSideSlideMechanisms1973}. All these contributions combined, we analyze the anomalous Hall coefficient, $R_s = \rhoAHE/M$, across different conductivity windows, given as follows:

\begin{equation}
    R_s =\frac{\rho_{\mathrm{AHE}}}{M}=\alpha\rho_{xx}+(\beta+b)\rho_{xx}^{2},
\label{eq:Rs}
\end{equation}

where the linear coefficient $\alpha$ captures the extrinsic skew scattering and the quadratic term combines the intrinsic ($\beta$) and the extrinsic side-jump ($b$) contributions~\cite{nagaosaAnomalousHallEffect2010,tianProperScalingAnomalous2009}.

\medskip

\textbf{\Fref{fig:ahe_overview}} shows how disorder modifies the AHE in our samples.
%the evolution of the anomalous Hall effect (AHE) in FeGe films as disorder increases from pristine (S0) to intermediate (S3) and highly irradiated (S5) levels.
In~\Fref{fig:ahe:temperature}, we plot the anomalous Hall resistivity, $\rho_{\mathrm{AHE,sat}}$, extracted from the saturation region ($\mu_0H>\SI{1.5}{\tesla}$), where skyrmions collapse into the ferromagnetic state and only conventional Hall contributions remain. All films exhibit a non-monotonic temperature dependence consistent with previous studies on clean FeGe~\cite{spencerHelicalMagneticStructure2018}, but disorder produces two salient changes: (i) the low-temperature AHE magnitude increases by a factor of \num{1.2} between S0 and S5, and, (ii) the broad peak near \SI{150}{\kelvin} shifts slightly higher. Scaling analyses and first-principles calculations agree that skew scattering, which grows linearly with impurity density, dominates the AHE once elastic mean-free paths fall below a few nanometres~\cite{adoAnomalousHallEffect2016, yuLargeAnomalousHall2025}.
Similar disorder-induced AHE enhancements have been observed in Fe$_3$GeTe$_2$~\cite{leeDisorderDrivenCrossover2025}, FePt~\cite{zimmermannInfluenceComplexDisorder2016}, and FeGe films~\cite{venutiInducingTunableSkyrmionantiskyrmion2024}.
%  and are often attributed to a crossover in dominant scattering processes (e.g., phonon vs. magnon scattering) or changes in the magnetization dynamics~\cite{}.
%where Ar or Ga$^{+}$ ions act as efficient skew centres.

\begin{figure*}[!h]
    \centering
    \begin{subcaptiongroup}
        \phantomcaption\label{fig:ahemech:cartoon}
        \phantomcaption\label{fig:ahemech:intrinsic}
        \phantomcaption\label{fig:ahemech:bad}
        \phantomcaption\label{fig:ahemech:skew}
        \phantomcaption\label{fig:ahemech:map}
    \end{subcaptiongroup}
    \includegraphics[width=\textwidth]{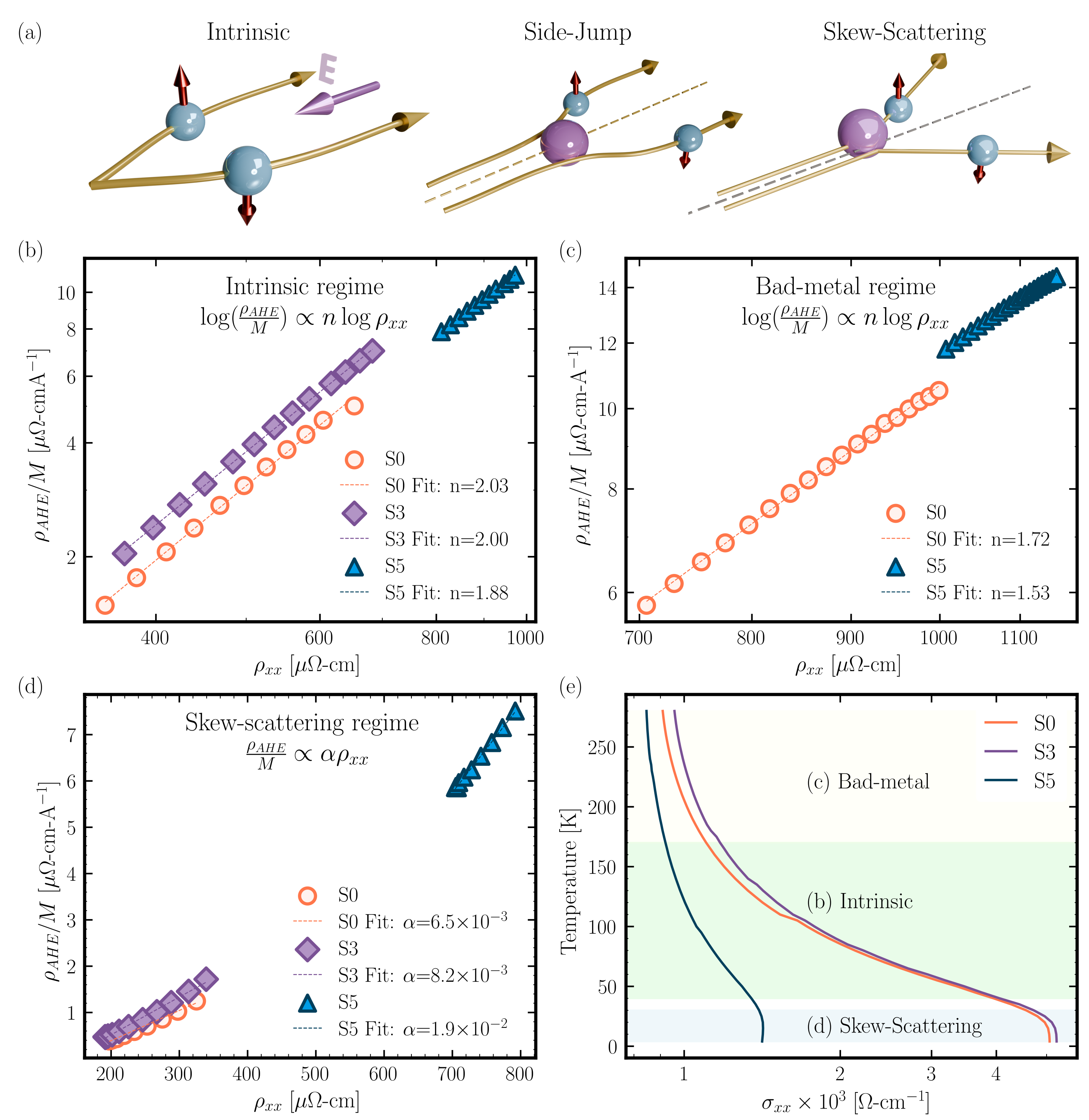}
    \caption{\protect\subref{fig:ahemech:cartoon} Schematic showing different AHE mechanisms. \protect\subref{fig:ahemech:intrinsic} Intrinsic window: $R_s\equiv\rho_{\mathrm{AHE}}/M$ scales nearly quadratically with $\rho_{xx}$; $R_s\propto\rho_{xx}^{n}$ with $n\approx2$ (fit exponents printed on panel). \protect\subref{fig:ahemech:bad} Bad-metal/mixed window: sub-quadratic power law $R_s=A\rho_{xx}^{\gamma}$ with $\gamma\approx1.5$–$1.7$, signaling deviation from the intrinsic limit at highest resistivities. \protect\subref{fig:ahemech:skew} Skew-scattering window: linear relation $R_s=\alpha\rho_{xx}$; extracted $\alpha$ increases systematically with disorder (S0 $<$ S3 $<$ S5; values shown). \protect\subref{fig:ahemech:map} Conductivity–temperature map: $\sigma_{xx}(T)$ for S0, S3, and S5, indicating the temperature ranges used for the three scaling analyses (shaded bands labeled to match panels). Symbols: S0 (orange circles), S3 (purple diamonds), S5 (blue crosses). Fits are performed over the highlighted $\rho_{xx}$ windows.}
    \label{fig:ahe_mech}
\end{figure*}

\medskip

To identify the dominant AHE mechanism at different defect densities, $\sigmaAHE$ is plotted against $\sigmaxx$ in \Fref{fig:ahe:sigmaxx}, with temperature as a parametric variable (color scale: purple at \SI{280}{\kelvin}, orange at \SI{10}{\kelvin}). This plot reveals that all samples traverse the canonical sequence of the three regimes identified by Nagaosa et al~\cite{nagaosaAnomalousHallEffect2010}. At the highest conductivities (low temperatures), the relationship between $\sigmaAHE$ and $\sigmaxx$ approaches linearity, indicating the dominance of extrinsic skew scattering, where impurities deflect carriers asymmetrically~\cite{smitSpontaneousHallEffect1958}. Increasing the temperature, thereby reducing $\sigmaxx$, horizontal plateaus are observed where $\sigmaAHE$ becomes nearly independent of $\sigmaxx$. This intrinsic regime is dominated by Berry curvature from the electronic band structure, which deflects carriers independent of scattering rate~\cite{vermaUnifiedTheoryAnomalous2022}. As temperature increases further, and the longitudinal conductivity drops, the trajectories bend downward into an intermediate regime where $\sigmaAHE \propto \sigmaxx^{\gamma}$ with $\gamma\approx1.6$, characteristic of competing intrinsic and extrinsic contributions~\cite{miyasatoCrossover2007, maryenkoObservationAnomalousHall2017, yangGiantUnconventionalAnomalous2020, }, also referred to as the bad-metal region.

\medskip

Overall, increasing the disorder shifts the entire trajectory toward lower values, pushing samples into extrinsic-dominated regimes at higher temperatures. This behavior parallels recent observations of giant AHE in spin-chirality-rich kagome metals, where enhanced impurity scattering amplifies the skew channel without significantly altering the intrinsic Berry curvature~\cite{fujishiroGiantAnomalousHall2021}. In what follows, we quantify the separate $\alpha$ and $(\beta+b)$ coefficients for different fluences and show how the same density of defects that stabilizes low-temperature skyrmions simultaneously tunes the balance between intrinsic and extrinsic anomalous Hall mechanisms.

\medskip

To extract scaling coefficients, we fit power laws, $A\rhoxx^n$, to the anomalous Hall coefficient $R_s \equiv \rhoAHE/M$ in appropriate resistivity windows~(\textbf{\Fref{fig:ahe_mech}}). \Fref{fig:ahemech:intrinsic} shows the intrinsic regime ($\SI{45}{\kelvin} < T < \SI{170}{\kelvin}$). Log-log plots yield $n= 2.03 \pm 0.05$ for S0, $n= 2.00 \pm 0.04$ for S3, and $n= 1.88 \pm 0.06$ for S5, confirming that Berry curvature dominates at intermediate conductivity where $R_s \propto \rhoxx^2$ ~\cite{tianProperScalingAnomalous2009}. Both intrinsic and side-jump mechanisms share this quadratic dependence~\cite{liAnomalousHallEffect2023}; previous FeGe studies suggest intrinsic contributions dominate in clean limits~\cite{porterScatteringMechanismsTextured2014, gallagherRobustZeroFieldSkyrmion2017}.

\medskip

At low conductivities with $T > \SI{170}{\kelvin}$, the exponent decreases: fitting $R_s \equiv A\rhoxx^\gamma$ yields $\gamma = 1.72 \pm 0.08$ for S0 and $\gamma = 1.53 \pm 0.07$ for S5, as shown in~\Fref{fig:ahemech:bad}. This crossover from quadratic ($n\approx 2$) to sub-quadratic ($\gamma \approx 1.5 - 1.7$) scaling indicates a transition from intrinsic-dominated to mixed intrinsic-extrinsic behavior, matching empirical universal dirty-metal, or bad-metal, scaling observed across ferromagnets.

\medskip

At highest conductivities with $T < \SI{70}{\kelvin}$, linear fits $R_s = \alpha\rhoxx$, extract the skew-coefficient, $\alpha$. We find $\alpha = 6.5 \pm 0.03 \times 10^{-3} \mu\Omega^{-1}\mathrm{cm}-A^{-1}$ for S0, increasing to $1.9 \pm 0.1 \times 10^{-2}$ for S5. This represents a threefold enhancement demonstrating that defects act as asymmetric scattering centers. \Fref{fig:ahemech:map} maps these regimes in temperature-conductivity space.
%: pristine S0 maintains intrinsic behavior to $\sim\SI{80}{\kelvin}$, where S5 enters mixed behavior by $\sim\SI{150}{\kelvin}$ and shows dominant skew scattering below $\SI{50}{\kelvin}$.
These results demonstrate that ion-irradiation provides systematic control over AHE mechanisms, with the same disorder that stabilizes low-temperature skyrmions also enhancing extrinsic electronic signatures.

\section{Conclusion}

Ion irradiation systematically controls both skyrmion stability and electronic transport in epitaxial FeGe films. Introducing disorder at fluences from $10^{11}$ to $10^{14}\,\,\mathrm{ions-cm}^{-2}$ produces two key effects. First, the topological Hall signal extends from a narrow window near \SI{200}{\kelvin} in pristine films down to \SI{4}{\kelvin} at the highest fluence, with peak THE amplitude more than doubling. This low-temperature stabilization is consistent with defects acting as trapping sites for skyrmions below their equilibrium phase boundary\cite{reichhardtStaticsDynamicsSkyrmions2022, hendersonSkyrmionAlignmentPinning2022}. Second, the anomalous Hall effect transitions from Berry-curvature-dominated ($\rhoAHE \propto \rhoxx^{2})$) to skew-scattering-dominated ($\rhoAHE \propto \rhoxx$) transport, with the skew coefficient increasing threefold. These parallel evolutions demonstrate that disorder modifies both real-space topological textures and momentum-space electronic scattering.

\medskip
The disorder-broadened low-temperature THE phase observed here has significant implications for understanding magnetic texture stability. In pristine FeGe, skyrmions exist only above approximately 80 K, whereas defect-rich samples exhibit robust THE signals down to 4 K. This extended phase space suggests that structural disorder fundamentally alters the energy landscape governing skyrmion formation and stability. However, the topological Hall effect can arise from several non-collinear magnetic structures beyond skyrmions, including helical spirals, magnetic bubbles, and other chiral domain configurations~\cite{kimbellChallengesIdentifyingChiral2022}. While skyrmions have been directly observed in FeGe via Lorentz transmission electron microscopy at high temperatures~\cite{yuRoomtemperatureFormationSkyrmion2011, yuSkyrmionFlowRoom2012}, the specific textures responsible for the\\low-temperature THE in irradiated films remain to be confirmed.

\medskip

Future work should directly image magnetic configurations in disordered regions to establish whether the enhanced low-temperature THE corresponds to stabilized skyrmions or alternative magnetic structures. Lorentz transmission electron microscopy can reveal real-space spin textures with nanometer resolution~\cite{yuRoomtemperatureFormationSkyrmion2011}, while magnetic force microscopy provides complementary surface sensitivity~\cite{heinzeSpontaneousAtomicscaleMagnetic2011}. Resonant X-ray scattering techniques can probe the helicity and periodicity of magnetic textures throughout the film thickness~\cite{christensen2024RoadmapMagnetic2024}. These measurements would distinguish true skyrmion lattices from disordered assemblies of magnetic bubbles or fragmented helical domains. Additionally, systematic current-driven dynamics measurements across disorder levels would quantify how defects modify skyrmion mobility and the skyrmion Hall angle~\cite{jiangDirectObservationSkyrmion2017}, establishing whether disorder-stabilized textures retain the ultralow critical currents characteristic of skyrmion motion. Such studies will determine whether disorder-broadened topological Hall phases represent extended skyrmion stability or the emergence of alternative chiral magnetic structures.

% =========================
% Experimental section
% =========================
\section{Experimental Section}\label{sec:methods}
% \todo[color=green!40]{Check all the numbers and conditions.}
\subsection{Sample Fabrication}
\label{sec:methods:fab}
Epitaxial B20-type FeGe films with a nominal thickness of \SI{300}{\nm} were grown on Si(111) using molecular beam epitaxy (MBE) at the Platform for Accelerated Realization, Analysis, and Discovery of Interface Materials (PARADIM) at Cornell University, following protocols similar to that used for $\textrm{Mn}_{x}\textrm{Fe}_{1-x}\textrm{Ge}$~\cite{turgutEngineeringDzyaloshinskiiMoriyaInteraction2018, venutiInducingTunableSkyrmionantiskyrmion2024, }. Films were ion-milled down to $\SI{80}{\nm}$ at the Center for Integrated Nanotechnologies (CINT) at Sandia National Laboratories. Final thickness was then confirmed by atomic force microscopy. Wafers were diced into \qtyproduct{4 x 4}{\milli\metre} pieces for magnetometry and \qtyproduct{6 x 6}{\milli\metre} chips for fabrication of Hall bars used in transport measurements. Six-terminal Hall bars were patterned using photolithography at the Washington Nanofabrication Facility (WNF) and etched using broad-beam Ar$^{+}$ ion milling at CINT.
% Each $6\times6$~mm$^{2}$ chip was solvent-cleaned (acetone, then IPA, N$_2$ dry), pre-baked at \SI{110}{\celsius} for \SI{60}{s}, spin-coated with AZ1512 at \SI{5000}{rpm}, and soft-baked at \SI{110}{\celsius} for \SI{60}{s}.\todo[inline, inlinewidth=5cm]{Add patterning details}. Ti/Au (\SI{5}{nm}/\SI{100}{nm}) was deposited by e-beam evaporation (FC-2000; \SI{0.5}{\angstrom\per\s} Ti, \SI{1.0}{\angstrom\per\s} Au) and lifted off in acetone for $\ge\,$\SI{2}{h}. Devices were then wire-bonded on chip carriers for low-noise four-terminal measurements.

\subsection{Ion-beam modification}
\label{sec:methods:irr}
Ion irradiation was performed using the NEC 3 MV Pelletron Tandem Accelerator at the Ion Beam Materials Laboratory in Los Alamos National Laboratory. Ne$^{++}$ ions of energy \SI{400}{\keV} were delivered at normal incidence with beam currents of $\sim\SI{18}{\nA}$ to systematically vary the defect density. Each chip was mounted on a Si backing wafer using double-sided carbon tape. The vacuum chamber was evacuated to a base pressure below\\$\SI{2e-6}{\torr}$ prior to irradiation. Five fluences: $10^{11},\,5\times10^{11},\,10^{12},\,10^{13},\,\SI[scientific-notation = true, print-unity-mantissa = false]{1e14}{\ions\per\square\cm}$, were applied to separate chips, with one unirradiated chip retained as a pristine reference (S0). Samples were labeled S1 through S5 in order of increasing fluence.

\medskip

Damage profiles were calculated using SRIM-2013 (Stopping and Range of Ions in Matter) in full-cascade mode with displacement energies of \SI{25}{\eV} for Fe and \SI{15}{\eV} for Ge~\cite{zieglerSRIMStoppingRange2010, Jain2013Commentary:Innovation}. Each simulation used $10^5$ ion trajectories. Displacements per atom (dpa) versus depth for different fluences are reported in the main text.

\subsection{Transport Measurements and Analysis}
\label{sec:methods:transport}
Hall resistivity measurements were performed in an Oxford Teslatron PT cryostat with a superconducting magnet providing out-of-plane magnetic fields $\mu_{0}H\in[-3,3]~\mathrm{T}$ and temperature control from \SIrange{4}{300}{\kelvin}. Samples were mounted on a non-magnetic leadless chip carrier. AC currents of $I_{\textrm{rms}}=\SI{20}{\uA}$ at frequencies $f_1 = \SI{39}{\hertz}$ (longitudinal) and $f_2=\SI{41}{\hertz}$ (transverse) were applied using internal current source of a Stanford Research Systems (SR) 860 lock-in amplifier. Longitudinal voltage $V_{xx}$ was measured with a SR830 lock-in amplifier, and transverse voltage $V_{xy}$ with a SR860 lock-in amplifier. A SR560 low-noise voltage preamplifier was used alongside the SR860, with a gain of $100$, and a low-pass filter with a cutoff of \SI{10}{\kilo\hertz} to suppress DC drift.

Resistivities were calculated from measured voltages using device geometry. For a Hall bar with a channel width $w$, length between voltage contacts $l$, and film thickness $t = \SI{80}{\nm}$, the longitudinal and transverse resistivities are:

\begin{equation}
    \rho_{xx} = \frac{V_{xx}}{I} \frac{w\cdot t}{\ell}, \quad \rho_{xy} = \frac{V_{xy}}{I} t.
\label{eq:resistivity}
\end{equation}

The total Hall resistivity $\rho_{xy}(H,T)$ contains three contributions:
\begin{equation}
\rho_{xy}(H,T) = R_0 H + \rho_{\mathrm{AHE}}(M(H,T)) + \rho_{\mathrm{THE}}(H,T),
\label{eq:hall_decomp}
\end{equation}
where $R_0$ is the ordinary Hall coefficient, $\rho_{\mathrm{AHE}}$ is the anomalous Hall resistivity proportional to magnetization $M$, and $\rho_{\mathrm{THE}}$ is the topological Hall contribution. These components were separated using the following procedure:

\begin{itemize}
    \item The ordinary Hall coefficient $R_0$ was determined from the high-field slope at $|\mu_0 H| > 2~\mathrm{T}$ where both $\rho_{\mathrm{THE}}$ and field-dependent $\rho_{\mathrm{AHE}}$ contributions are negligible. We fit $\rho_{xy}(H) = R_0 H + \rho_{\mathrm{offset}}$ to data averaged over field-up and field-down sweeps to minimize hysteresis effects. The offset $\rho_{\mathrm{offset}}$ accounts for contact misalignment.

    \item The anomalous Hall component was obtained by fitting the saturated high-field region $|\mu_0 H| > 1.5~\mathrm{T}$ to a linear relation $\rho_{\mathrm{AHE}} = R_s M(H,T)$ where $R_s$ is the anomalous Hall coefficient and $M(H,T)$ is the magnetization measured independently via SQUID magnetometry (see Methods~\Sref{sec:methods:mpms}). Magnetization data were interpolated to match transport temperature and field grids using cubic splines.
    
    \item The topological Hall resistivity was extracted as the residual:
        \begin{equation}
        \rho_{\mathrm{THE}}(H,T) = \rho_{xy}(H,T) - R_0 H - R_s M(H,T).
        \label{eq:the_extraction}
        \end{equation}
\end{itemize}

Field sweeps were performed at constant temperature between \SIrange{-3}{+3}{\tesla} in DRIVE mode.

\subsection{Magnetometry}
\label{sec:methods:mpms}
Magnetization measurements were performed on unpatterned \qtyproduct{4 x 4}{\milli\metre} chips using a Quantum Design MPMS3 superconducting quantum interference device (SQUID) magnetometer. Samples were mounted in low-magnetic-background drinking straw holders with a friction fit, with the film normal aligned perpendicular to the applied field. The sample volume used for converting the measured magnetic moment, $m$, to magnetization was calculated as $V = A \times t$, where $A=\SI{16}{mm\tothe{2}}$ is the normal chip area and $t=\SI{80}{\nm}$ is the film thickness.

\medskip

At each temperature, full magnetic moment hysteresis loops, $m(H)$, were collected from \SIrange{-3}{3}{\tesla} and back, recording data at stabilized field setpoints. Field step sizes were \SI{25}{\mT} for $|H|\le \SI{1.5}{\tesla}$ and \SI{50}{\mT} for $|H| > \SI{1.5}{\tesla}$ to capture rapid changes near phase transitions while maintaining reasonable acquisition items. The temperature grid spanned \SIrange{2}{10}{\kelvin} in \SI{2}{\kelvin} increments, followed by \SI{5}{\kelvin} steps from \SI{10}{\kelvin} to \SI{300}{\kelvin}. 

\medskip

The linear diamagnetic background from the Si substrate was removed by fitting $m_{\textrm{total}} = m_{\textrm{sample}}(H) + \chi_{Si}H$ to the high-field saturated regions ($|H| > \SI{2}{\tesla}$), where $\chi_{Si}$ is the substrate susceptibility. The corrected moment was then converted to volume magnetization: $M = m_{\textrm{sample}}/V$. Curie temperatures, $T_C$, were determined from the minimum in $dM/dT$, consistent with standard practice.

\subsection{STEM imaging and Multislice Electron Ptychography (MEP)}
\label{sec:methods:tem}
Cross-sectional FeGe/Si TEM specimens were prepared by a standard FIB lift-out process using a Thermo Fisher Helios G4 dual beam STEM/FIB. The final step of Ga-ion milling was done at \SI{5}{\kV}. STEM imaging was performed using an aberration-corrected Thermo Fisher Scientific Spectra 300 (S)TEM at \SI{300}{\kV}. The electron beam convergence half-angle was \SI{30}{\milli\radian}. MAADF collection angle range was \num{35} to \SI{67}{\milli\radian}. Scanning diffraction datasets for electron ptychography were collected with an EMPAD G2~\cite{philippVeryHighDynamicRange2022}, with a dwell time of \SI{100}{\us} and a maximum collection angle of about \SI{42}{\milli\radian}. MEP reconstruction is performed using the least-square maximum likelihood algorithm implemented in the \texttt{fold\_slice} package. The main reconstruction parameters are: 48 object slices, \SI{5}{\angstrom} slice thickness, scan step size \SI{0.5}{\angstrom}, and 8 probe modes.

\medskip
\medskip

\textbf{Supporting Information} \par %Please delete the Suppporting Information statement if it is not applicable. Please supply Supporting Information in another file. Supporting information should not be provided in .tex format
Supporting Information is available from the Wiley Online Library.

% Acknowledgements
\medskip
\textbf{Acknowledgments}  This material is based upon work supported by the National Science Foundation under grants DMR-1905909 (Serena Eley), DMR-2330562 (Serena Eley, C.G.), and DMR-2325089 (C.M.) at the Colorado School of Mines and the University of Washington. 
This work also made use of the Cornell Center for Materials Research shared
instrumentation facility and synthesis facilities at the Platform for the Accelerated Realization, Analysis, and Discovery of Interface Materials (PARADIM), which are supported by the National Science Foundation under Cooperative Agreement No. DMR-2039380. The Thermo Fisher Spectra 300 X-CFEG was
acquired with support from PARADIM, (NSF DMR-2039380), and Cornell University.
Part of this work was performed at the Center for Integrated Nanotechnologies, a DOE Office of Science User Facility. Sandia National Laboratories, managed and operated by NTESS, LLC, a wholly owned subsidiary of Honeywell International, Inc., for the U.S. DOE’s National Nuclear Security Administration. Los Alamos National Laboratory, an affirmative action equal opportunity employer, is managed by Triad National Security, LLC for the U.S. Department of Energy’s NNSA, under contract 89233218CNA000001. The views expressed in the article do not necessarily represent the views of the U.S. DOE or the United States Government.\\
This work was also partially supported by the National Science Foundation, Division of Materials Research under grant no. 2439947.\\
This research was also partially supported by NSF through the University of Washington Molecular Engineering Materials Center, a Materials Research Science and Engineering Center (DMR-2308979).\\
\medskip

\textbf{Author contributions.} S.E. (Serena Eley) conceived and designed the experiment.
H.P. grew the FeGe films.
Y.W. irradiated the samples.
C.G. and C.M. performed magnetization and transport studies, as well as data analysis.
S.E. (Sarah Edwards) and J.H.C. assisted with preliminary electrical transport measurements.
R.M.G. and T.M.L. assisted with Hall bar fabrication.
R.S. performed the ferromagnetic resonance measurements.
H.Y. performed STEM imaging and MEP with supervision from D.A.M..
C.G. wrote the manuscript, with contributions from C.M. and S.E. (Serena Eley). 
All authors commented on the manuscript.\\
\medskip

\textbf{Competing interests.} The authors declare no competing interests.\\
\medskip

\textbf{Data availability.} The data that support the findings of this study are available on Mendeley Data as a zip file. This includes .csv files containing the data used in figures, microscopy images (.tiff files), Python code used for the analysis, and an Origin file (.opju) containing all figures and data spreadsheets, which can be opened using Origin Viewer, a free application that permits viewing and copying of data contained in Origin project files.

% References
\medskip

% Use the following code if you wish to generate your bibliography with BibTeX;
% replace the string "MSP-template" below with the name(s) of
% the BibTeX data base(s) you want to use.
% The resulting bibliography-output (the content of the .bbl file)
% must be pasted back into this file before submission.
% Please also include your BibTeX data base file(s) in your submission
% so that we can re-run BibTeX if necessary.

\printbibliography

% Table of contents entry should be 50 - 60 words long
% Image should be 55 mm broad and 50 mm high or 110 mm broad and 20 mm high

% \begin{figure}
% \textbf{Table of Contents}\\
% \medskip
%   \includegraphics{toc-image.png}
%   \medskip
%   \caption*{ToC Entry}
% \end{figure}

\end{document}